\documentclass[showpacs,twocolumn,aps,prl,superscriptaddress,10pt]{revtex4-1}
\usepackage{amsmath,amssymb,amsfonts,bm,graphicx}
\usepackage{braket}

\usepackage{graphicx}
\usepackage{amsmath}
\usepackage{bm}
\usepackage{hyperref}
\hypersetup{%
	pdftitle={BaSh polaritons}, %
	pdfauthor={Zhiyuan Sun},%
	pdfpagemode={UseNone},%
	pdfstartview={FitH},%
	breaklinks=true,%
	citecolor=blue,%
	colorlinks=true,%
	linkcolor=blue,%
	urlcolor=blue
}

\usepackage[T1]{fontenc}
\usepackage{fourier}
\usepackage{baskervald}

\newcommand{\unit}[1]{\,\mathrm{#1}} 
\newcommand{\equa}[1]{Eq.~\eqref{#1}} %

\graphicspath{{./}}

\begin{document}

\title{Bardasis-Schrieffer polaritons in excitonic insulators}

\author{Zhiyuan Sun}
\affiliation{Department of Physics, Columbia University,
	538 West 120th Street, New York, New York 10027}

\author{Andrew J. Millis}
\affiliation{Department of Physics, Columbia University,
	538 West 120th Street, New York, New York 10027}
\affiliation{Center for Computational Quantum Physics, The Flatiron Institute, 162 5th Avenue, New York, New York 10010}
\date{\today}

\begin{abstract}
Bardasis-Schrieffer modes in superconductors are fluctuations in subdominant pairing channels, e.g., d-wave fluctuations in an s-wave superconductor. This Rapid Communication shows that these modes also generically occur in excitonic insulators.  In s-wave excitonic insulators, a p-wave Bardasis-Schrieffer mode exists below the gap energy, is optically active and hybridizes strongly with photons to form Bardasis-Schrieffer polaritons, which are observable in both far-field and near-field optical experiments. 
\end{abstract}

\maketitle

Sixty years ago, Bardasis and Schrieffer \cite{Bardasis1961} investigated  exciton-like sub-gap collective modes in superconductors produced by fluctuations in  channels different from the ground state, e.g., $d$-wave fluctuations in an $s$-wave superconductor. These modes are now referred to as Bardasis-Schrieffer modes (BaSh modes) \cite{Maiti2016,Allocca2019}. BaSh modes can be viewed as  collective  waves of transitions between different symmetry bound states of Cooper pairs.  In equilibrium superconductors BaSh modes typically do not couple linearly to long-wavelength radiation because a uniform electric field couples  to the center of mass motion, but not  the internal structure of a Cooper pair. A BaSh mode can couple to light in the presence of a supercurrent \cite{Allocca2019} or at nonzero momentum \cite{Sun2020}. Very recent Raman experiments have reported BaSh modes in iron based superconductors \cite{Kretzschmar2013,Bohm2014,Jost.2018}. In excitonic insulators \cite{Mott1961,Kozlov1965,Jerome1967, Kogar2017,Werdehausen2018}  the condensate is formed by  electron hole pairs. The opposite charge of the electron and hole means that a spatially uniform electric field may couple to the internal structure of a pair, thus  can excite, e.g., the $p$-wave BaSh mode in an $s$-wave condensate.

\begin{figure}
	\includegraphics[width=1.0\linewidth]{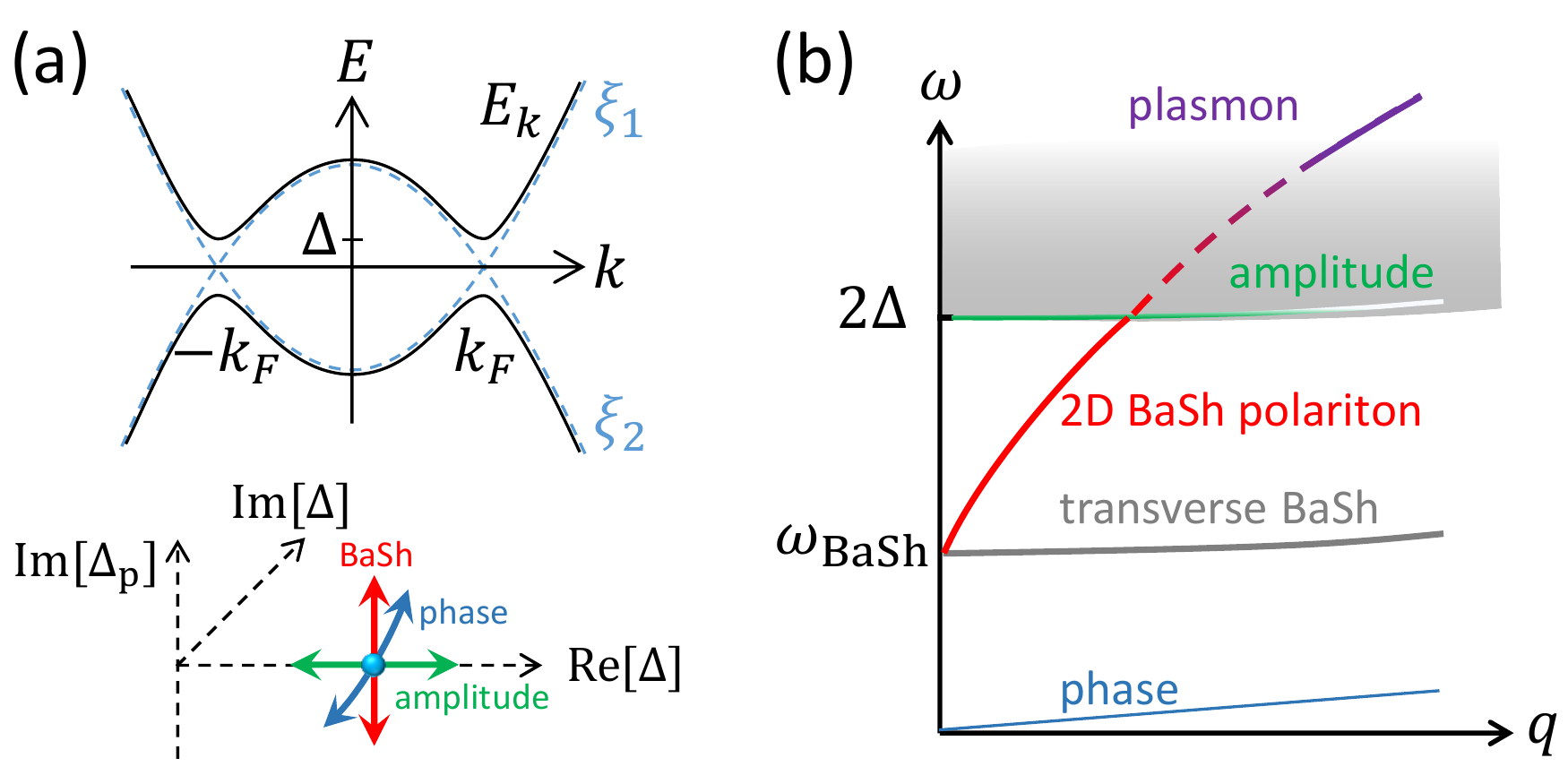}
	\caption{(a) Upper portion: noninteracting electron and hole bands (dashed lines) and renormalized bands (solid lines) in the excitonic insulating phase. Lower portion:  illustration of order parameter fluctuations  with axes being real and imaginary parts of the dominant order $\Delta$ and imaginary part of subdominant order ($\Delta_p$), showing amplitude, phase and BaSh modes. (b) Schematic frequency-momentum dispersion of the amplitude (green line) and phase (blue line) modes of an idealized two dimensional excitonic insulator  along with the transverse (grey line) and longitudinal (red line) BaSh modes. }
	\label{fig:schematic}
\end{figure}

In this Rapid Communication, we investigate the physics of Bardasis-Schrieffer modes in excitonic insulators using the minimal model 
\begin{align}
H =& \int {dr} \left[ \psi^\dagger
\left(
\xi(p-A) \sigma_3 + \phi 
\right)
\psi \right]
\notag\\
&  + 
\int {dr dr^\prime} V(r-r^\prime) \psi^\dagger(r) \psi(r) \psi^\dagger(r^\prime) \psi(r^\prime)
\,.
\label{eqn:hamiltonian}
\end{align}
Here $\psi^\dagger=(\psi_1^\dagger , \, \psi_2^\dagger)$ is the two component electron creation operator corresponding to the electron and hole bands labeled $1$ and $2$, $p=-i \hbar\nabla$, $\xi(p)=\varepsilon(p)-\mu$ is the kinetic energy, $(\phi,\, A)$ is the electromagnetic (EM) potential, $\sigma_i$ are the Pauli matrices in band space and we have set electron charge $e$ and speed of light $c$ to one. For notational simplicity, the EM field energy $F^{\mu\nu} F_{\mu\nu}/(16\pi)$ is not explicitly written in \equa{eqn:hamiltonian}. On the non-interacting level the numbers of electrons and holes are separately conserved and the electron and hole bands have the same dispersion but with opposite sign. We assume a negative gap, so that the two dispersions cross at a wavevector $k_F$ with fermi velocity $v_F$ as shown by the dashed lines in Fig.~\ref{fig:schematic}. In the two dimensional case of main interest here each band with mass $m$ contributes a carrier density $n/2=k_F^2/(4\pi)$ and a density of state $\nu/2=k_F/(2\pi \hbar v_F)$.  This model omits many features of real solids including any asymmetry between electron and hole bands,  coupling to phonons and the breaking of the idealized internal U(1) symmetry down to a discrete symmetry \cite{Mazza2019a}. These complications are not relevant to the basic physics we wish to consider here. 

$V(r-r^\prime)$ is  usually taken as the static limit of the screened Coulomb interaction;  in two dimensions (2D) $V(q)=\frac{2\pi}{q \epsilon(q)} \approx \frac{2\pi}{q +q_{\text{TF}}}$; within RPA  the Thomas-Fermi wave vector $q_{\text{TF}}=e^2 m/ \hbar^2$ does not depend on the carrier density. The $q$-dependence means that  higher angular momentum channels generically exist, so  BaSh modes are expected  in all excitonic insulators.

\emph{Ginzburg-Landau action---}Absorbing the intraband ($\sigma_3$-portion) of the interaction into the band gap and making a Hubbard-Stratonovich transformation of the partition function $Z=\int D[\bar{\psi},\psi] e^{-S}$ in the electron-hole pairing channel yields
\begin{align}
Z = & \int D[A] D[\bar{\psi},\psi] D[\bar{\Delta},\Delta]
e^{-S[\psi, A, \Delta]}
\end{align}
where the action
\begin{align}
S = \int d\tau dr 
\Bigg\{
&
\psi^\dagger 
G_{A\Delta}^{-1}
\psi
+ 
\sum_l \frac{1}{2 g_l}|\Delta_l|^2 
\Bigg\}
\label{eqn:HS_action}
\end{align}
describes coupled dynamics of the fermion field $\psi$, the EM field $A$ and the excitonic gap $\Delta_l$ with $l$ denoting the pairing channel (angular momentum in our case). The coupling constant $g_l$ in each channel is defined in the supplemental material \cite{SI}. 
The fermion kernel is 
\begin{align}
G_{A\Delta}^{-1}
=
\begin{pmatrix}
\partial_\tau  + \phi_1 + \xi(p-A_1) & \sum_l \Delta_l f_l(p) \\
\sum_l \bar{\Delta}_l \bar{f}_l(p)   &  \partial_\tau + \phi_2 - \xi(p-A_2)
\end{pmatrix}
\,
\label{eqn:gorkov}
\end{align}
where $f_l(p)$ is the pairing function in channel $l$. Since the electron and hole atomic orbitals can be at different spatial locations we have notated the possibility that they may feel different EM fields. In solid state realizations such as Ta$_2$NiSe$_5$ \cite{ Werdehausen2018,Kaneko.2013,Mazza2019a,andrich2020imaging} the orbitals are spatially close enough that both orbitals feel the same EM field; in electron-hole bilayers \cite{Fogler2014a,Calman2018,Eisenstein2014} the difference in fields may be important.

After integrating out the fermions, one obtains a Ginzburg-Landau action $S(\Delta,A)$. Its saddle point gives the mean field order parameters. We assume that the $l=0$ component of the interaction is the strongest and thus the ground state has $s$-wave pairing with mean field gap $\Delta=2\Lambda e^{-\frac{1}{g_s \nu}}$ which without loss of generality we set to be real. For simplicity we take the  pairing function $f_0=1$. The energy  cutoff $\Lambda$ depends on the interaction and is at the order of $v_F q_{\text{TF}}$ for screened Coulomb interaction \cite{Kozlov1965,Zittartz1967}.

The collective modes are fluctuations around the mean field configuration. We focus on the $p$-wave BaSh mode which couples to light already at zero momentum. 
The higher angular momentum BaSh modes, such as the $d$-wave one, are dark due to optical selection rule, and are closer to the $2\Delta$ gap in frequency due to typically weaker interactions in those channels. 
In $d$ dimensions the p-wave order parameter is a vector that transforms as a $d$ dimensional representation of the symmetry group ($O(d)$ neglecting lattice effects). Denoting the components of the p-wave gap by $\tilde{\Delta}_j$ we have
\begin{align}
\sum_l \Delta_l f_l(p)
=
\Delta+ R + i\Delta 2\theta + \sum_{j} \tilde{\Delta}_j f_j(p)
\,
\end{align}
where $f_j=k_j/k_F$ are the $p$-wave pairing functions and $j=x,y, ...$.
Here the fluctuations in the dominant order parameter have been explicitly separated into amplitude ($R$) and phase ($\theta$) degrees of freedom, while the p-wave fluctuations can be separated into real and imaginary parts as $\tilde{\Delta}_{j}=\Delta^{(1)}_{j}-i\Delta^{(2)}_{j}$. 

\emph{The BaSh mode action---}Expanding to quadratic order in the fluctuations around the mean field configuration, working in the gauge $\phi=0$, one obtains the effective action for the order parameter and EM field: 
\begin{align}
S(\Delta, A)= \frac{1}{2} 
\bigg(&
G^{-1}_R(q) R(-q) R(q) 
+ G^{-1}_\theta(q) \theta(-q) \theta(q) 
\notag\\
& + G^{-1}_{\text{BS}ij}(q) \Delta^{(2)}_{i}(-q) \Delta^{(2)}_{j}(q) +  K_{ij}(q) A_{i}(-q) A_j(q)
\notag\\
& + 2 C_{ij}(q) \Delta^{(2)}_{i}(-q) A_j(q)
\bigg)
\,
\label{eqn:s_gap_A}
\end{align}
where $q$ means both momentum and frequency and summation over $q$ and repeated indices is assumed. In the weak couping BCS regime only the  `imaginary' $p$-wave fluctuations $\Delta^{(2)}_{j}$ give rise to collective modes \cite{Sun2020} so we have not written the $\Delta^{(1)}_j$ terms here, but  briefly treat them in our discussion of the strong coupling (BEC) regime at the end and in the supplemental material \cite{SI}. 

$G_R$ and $G_\theta$ are the familiar amplitude and phase mode propagators. They are identical to those of a BCS superconductor \cite{Sun2020} due to the formal analogy of the action \equa{eqn:HS_action} to the BCS action, with the electron and hole band index mapped to the spin index in the superconductor.
The BaSh mode propagator 
\begin{align}
G_{\text{BS}ij}^{-1} = \frac{1}{g_p} \delta_{ij} + \chi_{\sigma_2 f_i,\,\sigma_2 f_j}= \left( \frac{1}{g_p} -\frac{1}{d g_s} - \frac{1}{d} \omega^2 F(\omega) \right) \delta_{ij}
\,
\label{eqn:G_BS}
\end{align}
is also identical to the superconducting case. The function $F$ describes the physics of quasiparticle excitations and is 
\begin{align}
F(\omega) = \sum_{k} \frac{1}{E_k(4E_k^2 - \omega^2)}
=
\frac{\nu}{4\Delta^2} \frac{2\Delta}{\omega} \frac{\mathrm{sin}^{-1}\left(\frac{\omega}{2\Delta}\right)}{\sqrt{1-\left(\frac{\omega}{2\Delta}\right)^2}}
\,
\end{align} 
which diverges as $1/\sqrt{2\Delta-\omega}$ as the frequency approaches the quasi particle excitation edge.

The key difference from superconductivity is the coupling to the EM field: the superconducting phase mode couples as $\partial_\mu\theta_{sc}\rightarrow \partial_\mu\theta_{sc}+A_\mu$, but in the excitonic case the neutrality of the particle-hole pair means there is no such coupling. On the other hand, the allowed dipole matrix element leads to the photon kernel
\begin{align}
K_{ij}(\omega) = \frac{n}{m} \delta_{ij} + \chi_{\sigma_3 v_i,\,\sigma_3 v_j} = \left(\frac{n}{m} - \frac{4}{d} v_F^2 \Delta^2 F(\omega) \right) \delta_{ij}
\,
\label{eqn:K_ij}
\end{align}
which contains pair breaking excitations described by $F(\omega)$ even without assistance of disorder.
Moreover, there is a linear coupling between the BaSh mode and the EM vector potential:
\begin{align}
C_{ij}(\omega)=\chi_{\sigma_2 f_i,\,\sigma_3 v_j}=-2i\Delta \omega \frac{v_F}{d} F(\omega) \delta_{ij}
\,.
\label{eqn:C_ij}
\end{align}

Due to $U(1)$ symmetries associated with the conservation of electron/hole numbers in the high temperature phase, the phase mode is gapless \cite{Jerome1967} with $G_\theta^{-1}=\nu\left( -\omega^2 + v_F^2 q^2/d \right)$ in the low energy limit. 
Lattice effects may reduce the $U(1)$ symmetry to a discrete one \cite{Mazza2019a} and open a gap to the phase mode dispersion. The amplitude mode has the gap $2\Delta$ and does not couple to light linearly at zero momentum. However, the BaSh mode couples to the electric field even at zero momentum because the latter exerts opposite forces on the electron and hole in an exciton, and excites it from the s bound state to p bound state. This induces a BaSh mode pole in the optical conductivity as we will show later.

\begin{figure}
	\includegraphics[width=0.8\linewidth]{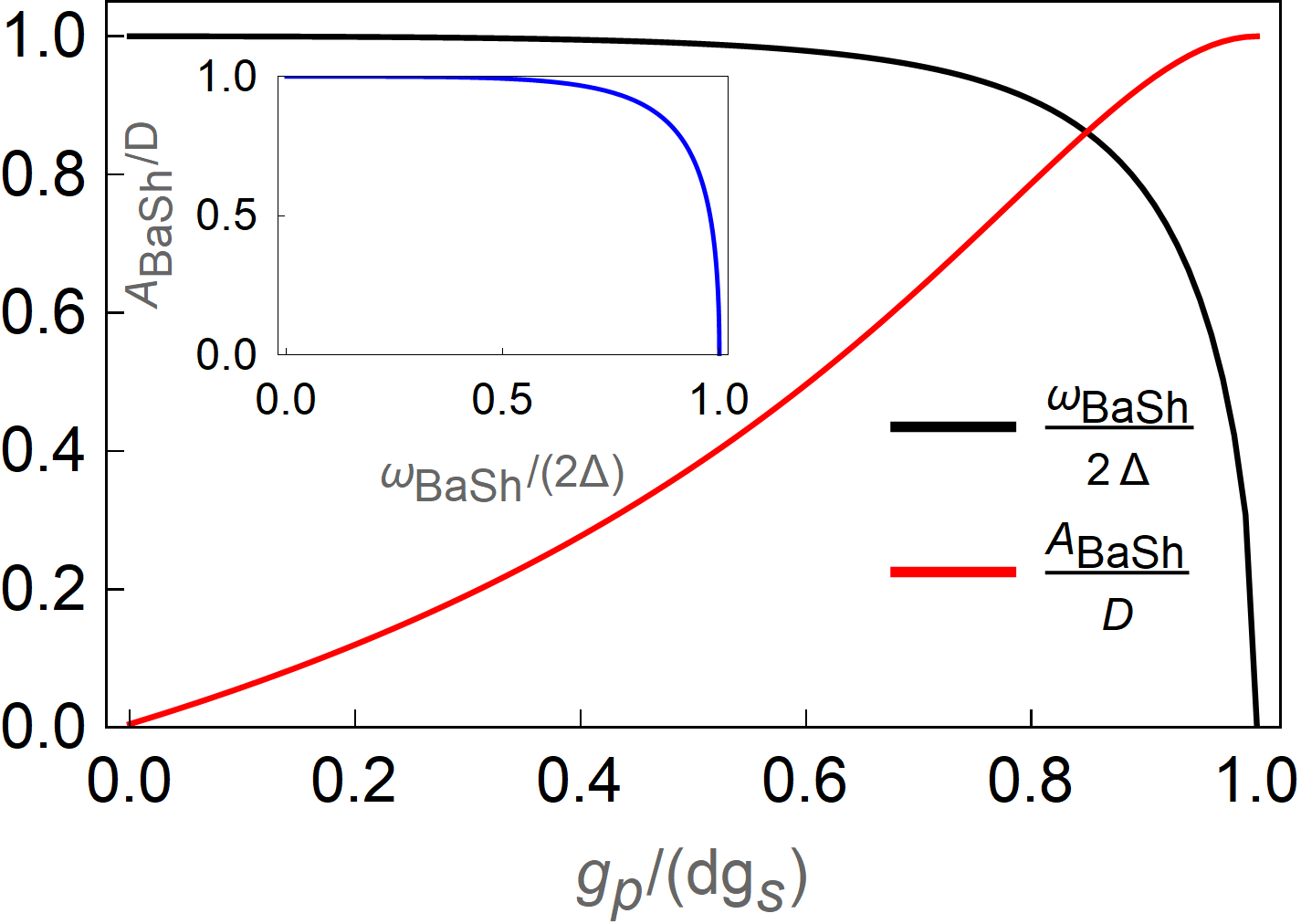}
	\caption{Black line: $q\rightarrow 0$  BaSh mode frequency as a function of $p$-wave coupling strength $g_p$ relative to $s$-wave coupling strength $g_s \nu=0.1$. Red line:  spectra weight of BaSh mode pole in the optical conductivity. Inset is the BaSh mode spectra weight as a universal function of BaSh mode frequency.}
	\label{fig:spectra_weight_BS}
\end{figure}

The root of \equa{eqn:G_BS} gives the BaSh mode frequency \cite{Bardasis1961,Maiti2015,Allocca2019,Sun2020} which decreases from $2\Delta$ to zero as $g_p$ grows from zero to $dg_s$, as shown in Fig.~\ref{fig:spectra_weight_BS}. In the weak and strong $p$-wave pairing limits, the BaSh mode frequencies are
\begin{align}
\omega_{\text{BaSh}}
=
2\Delta \left\{
\begin{array}{lc}
1- \frac{\pi^2}{8d^2} (\nu g_p)^2 
&  
\,(g_p \ll dg_s)
\\
\sqrt{ \frac{d}{ g_p \nu}- \frac{1}{ g_s \nu } }
&  \,(g_p \rightarrow dg_s)
\end{array}
\right. \,.
\label{eqn:sigma_n}
\end{align}
We define $\omega_{\text{BaSh}} \approx 2\Delta$ as the \emph{weak} BaSh mode case and $\omega_{\text{BaSh}} \ll 2\Delta$ as the \emph{strong} BaSh mode case. As $ g_p$ exceeds $d g_s$, the ground state order parameter starts to develop a $p$-wave component \cite{Maiti2015} and becomes an $s+ip$ state.


\begin{figure}
	\includegraphics[width=0.9\linewidth]{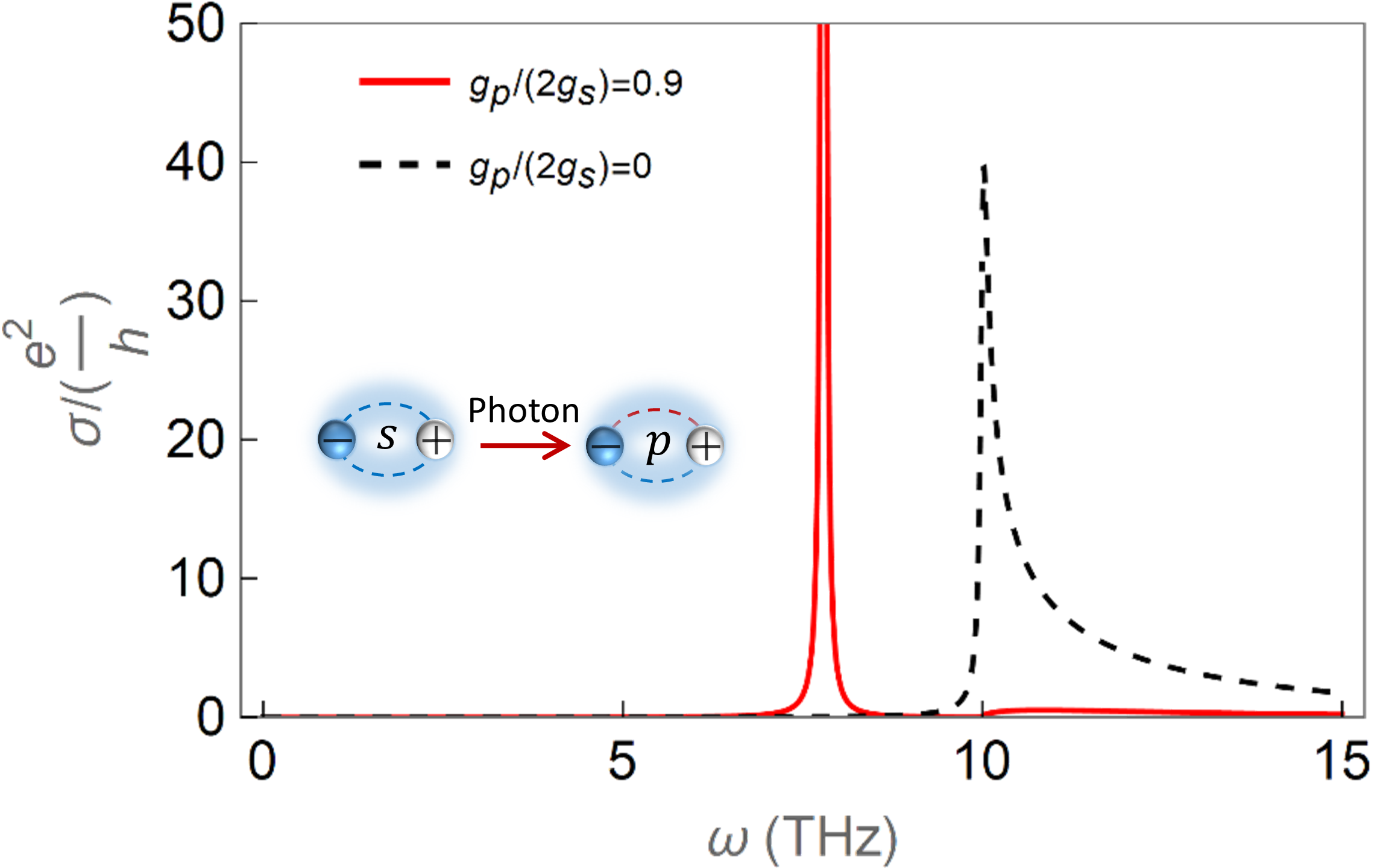}
	\caption{The real part of the optical conductivity of a 2D excitonic insulator with (red solid line) and without (dashed line) the BaSh mode contribution. The parameters are $\omega_{\text{BaSh}}=7.8 \unit{THz}$, $\Delta=5 \unit{THz}$, $n= 10^{12} \unit{cm^{-2}}$ and $v_F= 10^6 \unit{m/s}$. The line width of the BaSh mode is set to $\gamma_{\text{BaSh}}=0.01 \unit{THz}$. Inset: caricature of photon excitation process in which the photon excites the BaSh mode by converting $s$-excitons to $p$-excitons.   }
	\label{fig:sigma}
\end{figure}

\emph{Optical conductivity---}
Integrating out the order parameter fluctuations in $R$, $\theta$ and $\tilde{\Delta}_j$, one obtains the EM response kernel whose spatial part is the optical conductivity 
\begin{align}
\sigma(\omega)=\sigma_0 + \sigma_{\text{BaSh}}= \frac{i}{\omega} \left(\frac{n}{m} - \frac{4}{d} v_F^2 \Delta^2 F(\omega) \right) +\sigma_{\text{BaSh}}
\,.
\label{eqn:sigma}
\end{align}
We first consider $\sigma$ without the BaSh mode contribution $\sigma_{\text{BaSh}}$. In the zero frequency limit, the second term exactly cancels the first term such that the Drude spectral weight is zero, i.e., the system is an insulator \cite{Jerome1967}. The second term has zero total spectral weight since it decays faster than $1/\omega$ at large frequency and so acts to transfer the metallic phase Drude weight $D=\pi n/m$  to the above-gap  pair breaking excitations  in the excitonic insulating phase, as shown by the dashed line in Fig.~\ref{fig:sigma}.

The BaSh mode contribution 
\begin{align}
\sigma_{\text{BaSh}}(\omega) = \frac{4}{d^2} v_F^2 \Delta^2 F(\omega) F(-\omega) i\omega G_{\text{BS}}(\omega)
\,
\label{eqn:sigma_BS}
\end{align}
contains a pole below $2\Delta$. This term transfers spectral weight from the pair breaking excitations to the BaSh mode pole as shown by the red solid line in Fig.~\ref{fig:sigma}. In the BCS limit, the spectral weight
\begin{align}
A_{\text{BaSh}}(t)/D =  \frac{2 g(t)^2}{2g(t)+ t \partial_t g(t)}
\,
\label{eqn:BS_weight}
\end{align}
of the BaSh mode is a scaling  function of $t=\omega_{\text{BaSh}}/(2\Delta)$ where $g(t)=F(2\Delta t)4\Delta^2/\nu=\frac{1}{t} \frac{\mathrm{sin}^{-1}t}{\sqrt{1-t^2}} $. Starting from zero at $\omega_{\text{BaSh}}=2\Delta$, it grows until it reaches the total spectral weight  as $\omega_{\text{BaSh}} \rightarrow 0$, as shown by Fig.~\ref{fig:spectra_weight_BS}. In order for the BaSh mode frequency to be significantly below the gap, $g_p$ needs to be quite large  which also implies a very large BaSh mode spectra weight.

\emph{BaSh polariton---}There are two types of BaSh modes, which may be characterized as longitudinal (polarization parallel to momentum) and transverse (polarization perpendicular to momentum and $d-1$ fold degenerate). The longitudinal mode couples strongly to electromagnetic fluctuations, forming a BaSh polariton. In 2D, the polariton dispersion in the near field limit ($\omega \ll cq$) can be found from the zeros of the 2D dielectric function:
\begin{align}
\epsilon_{2D} = 1+ \frac{2\pi q i}{\omega} \sigma(\omega) =0
\,.
\label{eqn:dielectric}
\end{align}
Around zero momentum, the polariton frequency starts from $\omega_{\text{BaSh}}$ and shifts up linearly with momentum due to the Coulomb potential associated with the dipolar fluctuation. In the weak BaSh case, the polariton dispersion is 
\begin{align}
q = 
\left\{
\begin{array}{lc}
\frac{2^5}{\pi^2} \frac{\Delta}{D g_p \nu}  (\omega-\omega_{\text{BaSh}}) & \omega \rightarrow \omega_{\text{BaSh}}
\\ 
\frac{2 \Delta^2}{D
} \frac{1}{2/(g_p\nu)+1} 
& \omega \rightarrow 2\Delta
\end{array}
\right. 
\,.
\label{eqn:weak_BS_dispersion+}
\end{align}
Around zero momentum, the group velocity of the polariton is determined by the spectra weight of the BaSh mode pole: $v_g=\pi^2 g_p \nu D/(2^5\Delta)=\frac{\pi^2}{2^6} (g_p \nu) \frac{e^2}{\hbar v_F}  \frac{\varepsilon_F}{\Delta} v_F$ which is at the order of or larger than the fermi velocity if the fermi energy $\varepsilon_F\gg \Delta$. 
In the strong BaSh mode case, the optical conductivity \equa{eqn:sigma} becomes that of a Lorentzian oscillator: $\sigma \rightarrow \frac{D}{\pi} \frac{i\omega}{\omega^2-\omega_{\text{BaSh}}^2} $, and the BaSh polariton dispersion is simply
\begin{align}
q = 
\frac{1}{2D}(\omega^2-\omega_{\text{BaSh}}^2)
\,,
\label{eqn:strong_BS_dispersion}
\end{align}
just like the longitudinal phonon polaritons in 2D polar insulators \cite{Dai2019} and the exciton polaritons in 2D semiconductors in the near field regime (without a cavity).

\begin{figure}
	\includegraphics[width=0.8\linewidth]{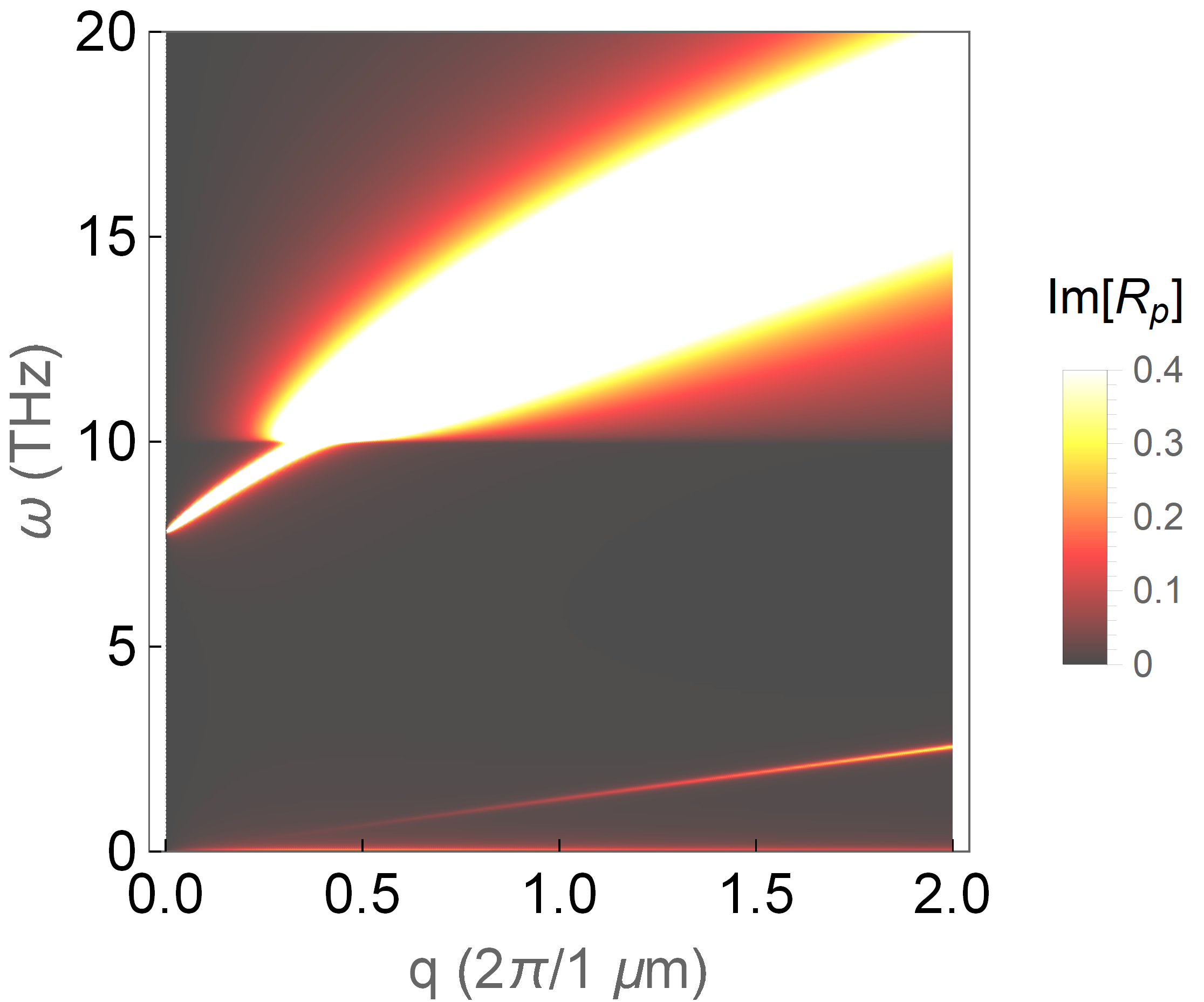}
	\caption{The near field reflection coefficient of a 2D excitonic insulator. The BaSh mode frequency is $\omega_{\text{BaSh}}=7.8 \unit{THz}$, corresponding to $g_s \nu =0.2$ and $g_p/(2 g_s) = 0.9$. The acoustic phase mode couples weakly to near field probe in the case of an electron hole bilayer with interlayer distance  $a=3 \unit{nm}$. The other parameters are $\Delta=5 \unit{THz}$, $n= 10^{12} \unit{cm^{-2}}$ and $v_F= 10^6 \unit{m/s}$. The damping line width of the BaSh mode is set to $\gamma_{\text{BaSh}}=0.02 \unit{THz}$ and that of the phase mode is $\gamma_{\text{phase}}=0.1 \unit{THz}$.}
	\label{fig:rp_BS}
\end{figure}

In the high frequency regime $\omega\gg 2\Delta$, the exciton physics becomes irrelevant and the optical conductivity approaches the Drude form $\sigma \rightarrow in/(m\omega)$, meaning that the BaSh polariton crosses over to the high energy plasmons.  The consequences for near field probes can be illustrated by the near field reflection coefficient~\cite{Basov2016,Low.2017,Sun2020}
\begin{align}
R_p(\omega,q) = 1-\frac{1}{\epsilon_{2D}(\omega,q)}
\,
\label{eqn:strong_BS_dispersion+}
\end{align}
plotted in Fig.~\ref{fig:rp_BS}.

The transverse BaSh mode does not couple to the coulomb interaction and is weakly dispersive: $\omega_q=\omega_{\text{BaSh}}+O(v_F^2 q^2/\Delta)$.
The separation of the transverse and longitudinal modes is similar to infrared active polar phonons \cite{Dai2019,Basov2016,Low.2017}. If the excitonic insulator is placed in an optical cavity similar to that studied in Ref.~\cite{Allocca2019}, the transverse BaSh mode can be red shifted due to coupling to a higher energy transverse photon. The combined photon/transverse BaSh mode is also referred to as a polariton.

In 3D, the bulk BaSh polariton frequency is determined by zeros of the 3D dielectric function $\epsilon_{3D}= 1+ \frac{4\pi i}{\omega} \sigma(\omega)$ which is typically too high in energy to be relevant. However, the transverse BaSh mode still has the dispersion shown in Fig.~\ref{fig:schematic}(b) and at zero momentum can be measured by far field optics. 

\emph{BEC---}In the strong coupling case, the excitons are strongly bound pairs and the transition to the excitonic condensate is essentially a Bose-Einstein condensation (BEC) of these preformed pairs. In the BEC state, the BaSh mode corresponds exactly to the atomic excitation of an $s$ bound state to a $p$ bound state, like a Hydrogen atom. The $1s\rightarrow 2p$ transition from light induced $s$ excitons has been observed by Merkl et al \cite{Merkl2019a}.   In this excitation, the `imaginary' and `real' $p$-wave order parameter fluctuations both appear,  corresponding to the interconversion of the dipole moment and current of an oscillating electric dipole. The BaSh mode frequency at zero momentum is thus the energy difference of the two bound states, i.e., $\hbar \omega_{\text{BaSh}}=(1-1/4)E_B$ in the case of Coulomb interaction where $E_B$=$\frac{1}{4}me^4/\hbar^2$ is the s state binding energy. Its spectra weight in the optical conductivity becomes $A_{\text{BaSh}}=\frac{3}{8}\frac{e^2 n_{\text{exciton}}}{m} c_p^2$ where $n_{\text{exciton}}$ is the number of excitons in the condensate. The dimensionless number $c_p\sim 1$ is defined as $\langle s|x|p\rangle=c_p a_{b}$ where $a_{b}=2\hbar^2/(me^2)$ is the Bohr radius.  

\emph{Electron hole bilayer---}In an electron hole bilayer, due to the non-negligible distance $a$ between the electron layer and the hole layer, the acoustic phase mode also couples to light since it is an exciton density fluctuation which induces local accumulation of z direction dipole moment. The resulting Coulomb potential shifts up the velocity of this `superfluid' sound.  To describe this mode, one needs to assume $A_1 \neq A_2$ in \equa{eqn:gorkov} to account for the difference of the EM field on the two layers.
Performing a local gauge transformation $(\psi_1, \psi_2) \rightarrow \left(\psi_1 e^{i\theta}, \psi_2 e^{-i\theta}\right)$ where $2\theta$ is the local phase of the $s$-wave gap \cite{Sun2020}, integrating out the fermions, one obtains the low energy effective Lagrangian 
\begin{align}
\mathcal{L} = -\frac{\nu}{2} \left( \partial_t \theta + \phi_a \right)^2 + \frac{n}{2m} \left(\nabla \theta - \mathbf{A}_a \right)^2 \,
\label{eqn:action_superfluid}
\end{align}
for the phase fluctuation where $(\phi_a,\, A_a)=(\phi_1-\phi_2,\, A_1-A_2)$ are the anti symmetric components of the EM field. The symmetric one does not couple to the phase mode. In the quasi static limit $\omega \ll cq$, the kinetic action of the anti symmetric EM field is just its electric field energy which  reads $S_{a}=\sum_q  \phi_a(q)^2/\left(2V_{\text{eff}}(q)\right)$ in the gauge $A_a=0$, with the mutually screened Coulomb kernel $V_{\text{eff}}(q)= (1-e^{-aq}) 2\pi/q$. Adding $S_{a}$ to \equa{eqn:action_superfluid} and solving the equation of motion, one obtains the dispersion of the phase mode
\begin{align}
\omega_{\text{phase}}(q) = q\sqrt{\left(\frac{1}{\nu} + V_{\text{eff}}(q) \right) \frac{n}{m} }  \approx q \sqrt{v_F^2/d + 2Da} \,
\label{eqn:phase_mode_dispersion}
\end{align}
which is the same as the anti symmetric plasmon mode of double layer superconductors \cite{Sun2020}. 
A nonzero tunneling between the layers induces a Josephson effect in the electron hole bilayer system \cite{Fogler2001} and gives a nonzero gap to the phase mode. But we don't consider this physics here.

The response of the phase mode to near field probe can be represented by its contribution to the near field reflection coefficient \cite{Sun2020}
\begin{align}
R_{\text{phase}}(\omega,q) = \frac{1}{4\pi} \frac{\frac{n}{m} q^3 V_{\text{eff}}(q)^2}{\omega^2- \omega_{\text{phase}}(q)^2} 
\,.
\label{eqn:rp_phase}
\end{align}
In the case of $2Da \gg v_F^2$ \cite{Fogler2014a,Calman2018,Eisenstein2014}, the phase mode shows up in the near field response with a spectra weight of $(aq)^2 \sqrt{n/(ma)} (2\pi )^{3/2}/8$, smaller than that of the BaSh polariton by roughly the factor $(aq)^{3/2} \ll 1$.

\emph{Discussion---}We introduced a class of collective modes to excitonic insulators: the BaSh polaritons. Our work bridges the area of excitonic insulators/exciton condensates \cite{Fogler2014a,Calman2018,Eisenstein2014, xue2020higgs,Nandkishore.2010,Li2017,Kogar2017,
Werdehausen2018,Kaneko.2013,Mazza2019a,andrich2020imaging} with the field of near field optics \cite{Lundeberg2017,Basov2016,Low.2017,Ni2018a} and will stimulate new classes of experiments and theoretical studies of photo induced nonequilibrium dynamics of excitonic insulators, and its effects on photo current/high harmonic generation. As a low loss (sub gap) propagating wave which can be easily excited by photons, the Bash polariton is a promising information carrier in nano optical devices. 

In electron hole bilayers made of transition metal dichalcogenides (TMD) \cite{Fogler2014a,Calman2018}, semiconductor quantum wells \cite{Eisenstein2014, xue2020higgs}, bilayer \cite{Nandkishore.2010} and double bilayer graphene \cite{Li2017}, the BaSh mode is the only optically active collective mode at energies close to the gap. In solid state excitonic insulator candidates, such as Ta$_2$NiSe$_5$ \cite{Werdehausen2018,Kaneko.2013,Mazza2019a,andrich2020imaging,Lu2017,Seo2018}, $1T$-TiSe$_2$ \cite{Kogar2017} and possibly nodal-line semi metals \cite{Shao.2020},  lattice effects complicate the interpretation, but BaSh modes are still expected to be observable, which can be predicted by our RPA type formalism applied to the specific interaction and band structure there. In all of these systems,  far field optics is a powerful  probe of the transverse BaSh mode and near field optics \cite{Lundeberg2017,Basov2016,Low.2017,Ni2018a} is the ideal tool to probe BaSh polaritons. 


In order for the BaSh mode to be well separated from the excitation continuum, one needs a substantial relative $p$-wave interaction to the $s$-wave one (Fig.~\ref{fig:spectra_weight_BS}). This can be realized by, e.g., screened Coulomb interaction in high carrier density ($\sim 10^{13} \unit{cm}^{-2}$) electron hole bilayers on high dielectric substrates (supplemental material \cite{SI}).
The linewidth of the BaSh mode is also an experimentally important issue.  At low temperature, electronic contributions are suppressed by the quasiparticle gap, but the mode may be broadened by inhomogenous broadening from disorder \cite{Zittartz1967}, and by decaying into phonons and other modes. As the temperature is increased, thermally excited carriers will play an increasingly important role in damping the BaSh modes which is an issue for future research. 

\begin{acknowledgements}
	We acknowledge support from  the Department of Energy under Grant DE-SC0018218. We thank T. Kaneko, D. Golez and W. Yang for helpful discussions.
\end{acknowledgements}

\bibliographystyle{apsrev4-1}
\bibliography{./Excitonic_Insulator}

\pagebreak
\widetext
\begin{center}
	\textbf{\large Supplemental Material for `Bardasis-Schrieffer polaritons in excitonic insulators'}
\end{center}
\setcounter{equation}{0}
\setcounter{figure}{0}
\setcounter{table}{0}
\setcounter{page}{1}
\makeatletter
\renewcommand{\theequation}{S\arabic{equation}}
\renewcommand{\thefigure}{S\arabic{figure}}
\renewcommand{\bibnumfmt}[1]{[S#1]}
\renewcommand{\citenumfont}[1]{S#1}

\section{The strength of pairing channels}
\label{apendix:gs_gp}
In the BCS regime of two dimensional excitonic insulators, assuming that the excitonic effects occur near a high symmetry point so lattice effects are unimportant, we can choose $f_l=\cos (l\theta_k)$ or $\sin (l\theta_k)$ and the corresponding pairing interaction is $g_l= \frac{1}{2\pi} \int d\theta \cos(l\theta) V\left(2 k_F \sin(\theta/2) \right)$. Note that for $l=0$, the $1/2\pi$ factor should be changed to $1/4\pi$. For Thomas-Fermi screened interaction  $V(q)= \frac{2\pi}{\epsilon (q +q_{\text{TF}})}$ in 2D where $q_{TF}/(2k_F)=\alpha=e^2/(\epsilon \hbar v_F)$ and $\epsilon$ is the dielectric constant of the environment, the $s$-wave pairing strength is 
\begin{align}
\nu g_s&=\nu \frac{1}{4\pi} \int d\theta \frac{2\pi}{2 k_F |\sin(\theta/2)| + q_{TF} } 
=\frac{\alpha}{\sqrt{1-\alpha^2}}
\frac{1}{\pi}
\mathrm{Tanh}^{-1}\left( \sqrt{1-\alpha^2} \right)
\label{eqn:g_s}
\end{align} 
and the $p$-wave one is
\begin{align}
\nu g_p&=\nu \frac{1}{2\pi} \int d\theta \frac{2\pi \cos\theta}{2 k_F |\sin(\theta/2)| + q_{TF} } 
\notag\\
&= \alpha
\Bigg[-\frac{4}{\pi} + 2 \alpha + \frac{4}{\pi}
\frac{\left(1-2\alpha^2\right)}{\sqrt{1-\alpha^2}} 
\Bigg( 
\mathrm{Tanh}^{-1} \left(\sqrt{1-\alpha^2} \right)
- \mathrm{Tanh}^{-1}\left(\frac{\sqrt{1-\alpha^2}}{1+\alpha} \right)
\Bigg) \Bigg]
\label{eqn:g_p}
\end{align} 
where $\nu=k_F/(\pi \hbar v_F)$ is the normal state density of state without spin degeneracy and $\alpha=e^2/(\epsilon \hbar v_F)$ is the `fine structure constant' in this system. 

\begin{figure}
	\includegraphics[width= 0.8\linewidth]{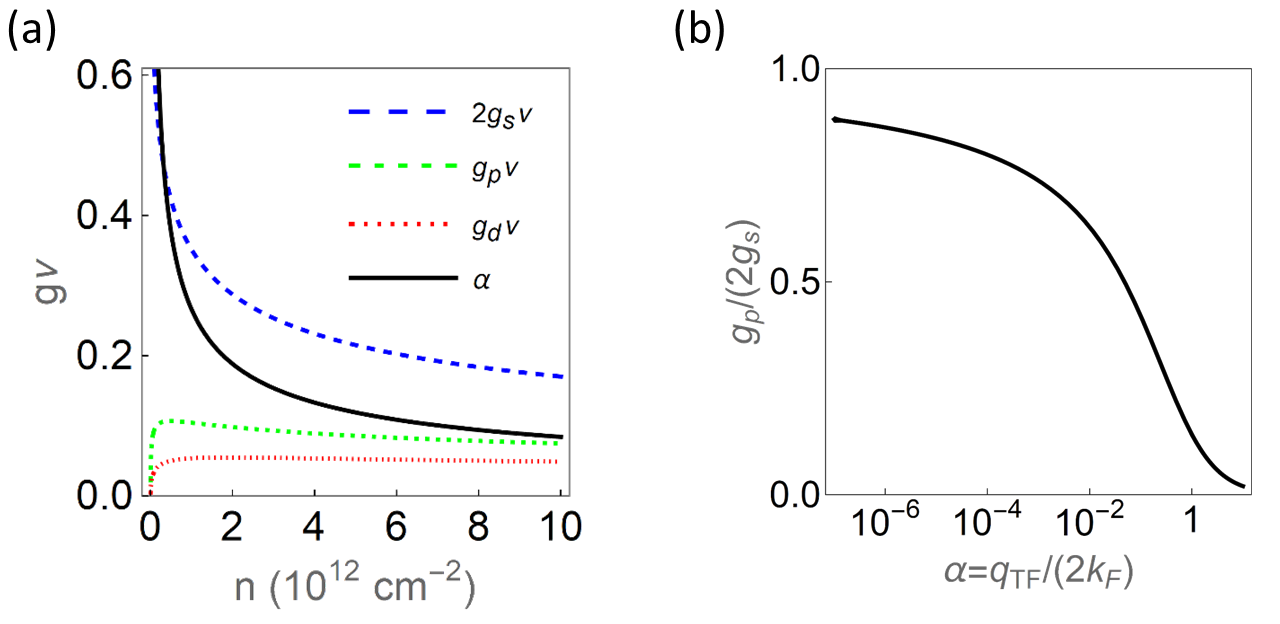}
	\caption{(a) The $s$,$p$,$d$-wave components of the screened Coulomb interaction in 2D and the `fine structure constant' $\alpha=e^2/(\epsilon \hbar v_F) = q_{TF}/k_F$ as functions of electron density $n_i=m^2 v_F^2/(4\pi\hbar^2)$ computed from Eqs.~\eqref{eqn:g_s} and \eqref{eqn:g_p} using $m=0.05 m_e$ and $\epsilon=10$. (b) The ratio $g_p/(2g_s)$ as a function of $\alpha=q_{TF}/(2k_F)$.  For $\alpha \ll 1$, i.e., in the high density case, $g_p/(2g_s)$ becomes considerable and approaches one in the high density limit.  Spin degeneracy is neglected.}
	\label{fig:gs_gp}
\end{figure}

The pairing interactions are shown in Fig.~\ref{fig:gs_gp} for the screened Coulomb interaction in 2D.
To obtain a substantial $g_p/(2 g_s)$, one needs the high density case where the fermi velocity is large so that the Thomas fermi wave vector is smaller than the fermi momentum: $q_{TF}/(2k_F)=\alpha=e^2/(\epsilon \hbar v_F) \ll 1$. Stronger dielectric screening of the environment can further reduce $\alpha$ and increase $g_p/(2 g_s)$. Moreover, a non-negligible interlayer distance $a$ changes the bare electron-hole Coulomb attraction into $V(r)=1/\sqrt{r^2+a^2}$, making it more nonlocal and thus can lead to a larger  $g_p/(2 g_s)$.

\section{Correlation functions}
\label{apendix:correlation_functions}
The correlation function $\chi_{\sigma_i \sigma_j}$ is defined as
\begin{align}
\chi_{\sigma_i \sigma_j}(q) = 
\left\langle \hat{T} 
\left(\psi^\dagger \sigma_i \psi \right)_{(r,t)}
\left(\psi^\dagger \sigma_j \psi \right)_0  
\right\rangle \bigg|_q 
=
\sum_{\omega_n, k}
Tr\left[ G(k,i\omega_n) \sigma_i G(k+q,i(\omega_n+\Omega))  \sigma_j \right]
\,
\label{eqn:chi_defi}
\end{align}
where $\hat{T}$ is the time order symbol, $x=(\mathbf{r},t)$, $q=(\mathbf{q},i\Omega)$ and
\begin{align}
G(k,i\omega_n) &= G_\Delta(k,i\omega_n) = 
\left\langle \hat{T} 
\psi (x)
\psi^\dagger(0) 
\right\rangle \bigg|_{k,i\omega_n}
=\frac{1}{i\omega_n - \xi_k \sigma_3 -\Delta\sigma_1 }
\,
\end{align}
is the electron Green's function. The BaSh mode propagator is
\begin{align}
G_{\text{BS}xx}^{-1} = \frac{1}{g_p} + \chi_{\sigma_2 f_x,\,\sigma_2 f_x}(\omega) 
= \frac{1}{g_p} + \sum_{k}
\frac{4 \cos^2(\theta_k) E_k }{\omega^2 - 4E_k^2}
= \frac{1}{g_p} - \frac{1}{d} \left( \frac{1}{g_s} +\omega^2 F(\omega) \right)
\,,
\label{eqn:BS_G_detail}
\end{align}
where the last equality comes from the gap equation $\frac{1}{g_s}=\sum_{k}
\frac{1}{E_k}$. 
The photon kernel is
\begin{align}
K_{xx}(\omega) = \frac{n}{m} + \chi_{\sigma_3 v_x,\,\sigma_3 v_x}(\omega) 
= 
\frac{n}{m} + \frac{1}{d}
\sum_{k}
\frac{\Delta^2}{E}
\frac{4 v_F^2}{\omega^2 - 4E^2}
= \frac{n}{m} - \frac{4}{d} v_F^2 \Delta^2 F(\omega)
\,.
\label{eqn:photon_kernel}
\end{align}
The linear coupling between BaSh mode and the EM vector potential is
\begin{align}
C_{ij}(\omega)=\chi_{\sigma_2 f_i,\,\sigma_3 v_j}(\omega)=
i \Delta \omega \sum_{k}
\frac{2 }{E_k}
\frac{f_i(k) v_j(k)}{\omega^2 - 4E^2}
=
-2i\Delta \omega \frac{v_F}{d} F(\omega) \delta_{ij}
\,.
\label{eqn:C_ij_detail}
\end{align}

\section{Strong coupling case}
The full quadratic action for the two components of the $p$-wave fluctuations is
\begin{align}
S_j(\tilde{\Delta})= 
\frac{1}{2}\sum_q  
\begin{pmatrix}
\Delta^{(1)}_{j} & \Delta^{(2)}_{j}
\end{pmatrix}_{-\omega}
\hat{M}
\begin{pmatrix}
\Delta^{(1)}_{j} \\
\Delta^{(2)}_{j}
\end{pmatrix}_q
=
\frac{1}{2}\sum_q& 
\begin{pmatrix}
\Delta^{(1)}_{j} & \Delta^{(2)}_{j}
\end{pmatrix}_{-q}
\begin{pmatrix}
\frac{1}{g_p} +\chi_{\sigma_1 f_j,\,\sigma_1 f_j}(q)  & \chi_{\sigma_1 f_j,\,\sigma_2 f_j}(q) \\
\chi_{\sigma_2 f_j,\,\sigma_1 f_j}(q) & 
\frac{1}{g_p} +\chi_{\sigma_2 f_j,\,\sigma_2 f_j}(q)
\end{pmatrix}_{q}
\begin{pmatrix}
\Delta^{(1)}_{j} \\
\Delta^{(2)}_{j}
\end{pmatrix}_q
\,
\label{eqn:s_real_imaginary}
\end{align}
which when restricted to zero momentum fluctuations simplifies to
\begin{align}
S_j(\tilde{\Delta})=
\frac{1}{2}\sum_\omega& 
\begin{pmatrix}
\Delta^{(1)}_{j} & \Delta^{(2)}_{j}
\end{pmatrix}_{-\omega}
\begin{pmatrix}
\frac{1}{g_p} - \sum_{k}  \frac{f_j(k)^2}{4E_k^2-\omega^2} \frac{4\xi_k^2}{E_k} & 
- 2i\omega \sum_{k}  \frac{f_j(k)^2}{4E_k^2-\omega^2}  \frac{\xi_k}{E_k} 
\\
2i\omega \sum_{k}  \frac{f_j(k)^2}{4E_k^2-\omega^2} \frac{\xi_k}{E_k} & 
\frac{1}{g_p} - \sum_{k}  \frac{f_j(k)^2}{4E_k^2-\omega^2} 4 E_k
\end{pmatrix}_{\omega}
\begin{pmatrix}
\Delta^{(1)}_{j} \\
\Delta^{(2)}_{j}
\end{pmatrix}_\omega
\,.
\label{eqn:s_real_imaginary_omega}
\end{align}
The collective mode frequencies are determined by the zeros of the determinant of the matrix in \equa{eqn:s_real_imaginary_omega}. 

In the weak coupling (BCS) limit studied in the main text,  the factor $\xi_k/E_k$ changes sign as $k$ crosses $k_F$ so that in the off-diagonal term the sum of $k$ gives a small value,     $\mathcal{O}(\Delta/\mu)$ relative to the diagonal terms, and in  the $\Delta^{(1)}-\Delta^{(1)}$ term the factor $\xi_k^2/E_k^2$ ensures that the $\sum_{k}  \frac{f_j(k)^2}{4E_k^2-\omega^2} \frac{4\xi_k^2}{E_k}$ does not diverge as $|\omega|\rightarrow 2\Delta$, so there is no zero of the inverse response function associated with the real part of $\Delta_j$.  The $\Delta^{(2)}-\Delta^{(2)}$ term is just the BaSh kernel studied in the main text. 

Away from the weak coupling BCS regime, the off-diagonal terms become non-negligible which
means the real and imaginary fluctuations are mixed together in the BaSh mode and some details of the structure of the individual terms change.  However, we find that the determinant of the BaSh mode matrix 
still has one root at frequencies less than the gap; this root is at a frequency lower than the $\omega_{\text{BaSh}}$ defined
in Eq.~(7) of the main text, meaning the BaSh mode frequency is pushed down by this cross coupling, and the eigenvector
of this mode is thus of mixed imaginary-real characters. The appearance of the BaSh mode in optical
conductivity stays qualitatively the same.

\end{document}